\documentstyle[preprint,prl,aps,epsf]{revtex}   
\def\lsim{\raise0.3ex\hbox{$\;<$\kern-0.75em\raise-1.1ex\hbox{$\sim\;$}}}
\def\gsim{\raise0.3ex\hbox{$\;>$\kern-0.75em\raise-1.1ex\hbox{$\sim\;$}}}

\def\kpc{\,{\rm kpc}}

\def\km{\,{\rm km}}
\def\s{\,{\rm s}}

\def\kmps{\km\,\s^{-1}}

\def\dm{_{\rm DM}}
\def\rmsvel{\langle v^2\rangle^{1/2}}
\def\rmsveldm{\langle v^2\rangle_{\dm}^{1/2}}
\def\rmsveldmsun{\langle v^2\rangle_{\dm,\odot}^{1/2}}

\def\vc{V_c}
\def\vcsun{V_{c,\odot}}
\def\vcinf{V_{c,\infty}}

\title{ Reply to Comment on "Dispersion Velocity of Galactic
 Dark Matter Particles" by Gates et al.}

\author{ 
R. Cowsik\footnote{cowsik, charu, pijush@iiap.ernet.in}, 
Charu Ratnam and P. Bhattacharjee}

\address{Indian Institute of Astrophysics, Bangalore 560 034, INDIA.}

\begin{document}
\maketitle
\centerline{(Received:\hspace{5cm})}

\newpage
\tighten
Amongst the various issues raised by Gates et al (GKT)\cite{gkt} 
the most crucial one is their claim
that our result disagrees with the observations; this is not true. 
Their claim stems from (i) confusion of model-dependent results 
(obtained on the basis of certain mass models) with actual 
observational constraints, 
(ii) confusing the classical mass models of the halo with models
that probe its phase space structure, and (iii) using the notion of 
superposition
not allowed by the self-consistent Boltzmann-Poisson equations which 
involve a 
non-linear coupling among the various components. Indeed, the very 
purpose
of our Letter\cite{crb} was to present a method which would 
sensitively probe the
density and dispersion velocity of dark-matter particles in the 
solar 
neighborhood, circumventing some of the problems encountered in 
previous analyses.

Let us first address the claim that the velocity dispersion of the 
dark 
matter particles in the solar neighborhood, 
$\rmsveldmsun = 270 \kmps$, based on the 
formula $\rmsveldm=\sqrt{\frac{3}{2}}\,\vcinf$. 
For this claim to be valid,
two conditions need to be satisfied: (1) The dynamics has to be 
that of a 
single component isothermal sphere so that the formula is 
applicable, and 
(2) the asymptotic circular speed, $V_{c,\infty}$, should be 
$\sim220\kmps$. The first of
these conditions is violated in the problem at hand; in the central 
regions
of the galaxy the density of the visible matter exceeds that of the 
dark matter
by factors $\sim 1000$. Even the integrated mass of the dark matter 
within a 
sphere of radius $R_\odot\sim 8.5\kpc$ is smaller than that of the 
visible matter.
Thus the above asymptotic relation between $\rmsveldm$ and $\vc$ is 
established only
at much larger distances, as we discuss in our response \cite{crb2} 
to the 
comment by Evans\cite{evans}. Secondly, there is no observational 
basis for 
the claim that $\vcinf=\vcsun = 220\kmps$. After extensive review 
Fich and Tremaine (cited in \cite{crb}) concluded that the rotation 
curve 
continues to rise beyond $R_\odot$. Indeed all available rotation 
curve data 
up to $R\sim20\kpc$ have been incorporated into Fig. 1 of 
\cite{crb} which forms the basis of our results.  

The observations of halo stars and globular clusters are also not 
in conflict
with our results. Frenk and White, as well as Norris and Hawkins   
(Ref. 2 of GKT), 
have discussed both the limitations and the uncertainties 
involved in the analysis of the problem. 
It is to be emphasised that all previous analyses of the problem, 
including
those in Ref.2 of GKT, were concerned with the {\it
mass distribution} in the halo, rather than its  
phase space structure. Even for the simpler problem of determining 
the density distribution, 
it is necessary to measure the six variables ($\vec{r}$,$\vec{v}$) 
for each of
the objects under study, and furthermore, a large statistical sample 
of these objects is needed. 
GKT's comment\cite{gkt} seems to foster the impression 
that the results of our analysis are inconsistent with observation. 
However, 
the so-called observations are in fact no more than model-dependent 
inferences.
This is clear from the fact that out of the six variables only four, 
namely,   
($\vec{r}$, $v_{r}$), 
have been measured, leaving the models highly underconstrained. 
Also contrary
to the assertion of GKT, inspection of Fig. 5 of Frenk and White 
shows that the parameters of the halo mass models have a wide 
dispersion, but
they are not inconsistent with our results. In this context, we do 
not know how GKT obtained the
value of $\rmsveldmsun\sim200\kmps$, as this value does not appear 
in any of the papers cited in Ref. 2 of GKT. 

The analysis by Gates et al (Ref. 3 of GKT) 
also does not probe the phase space structure of the halo. Different 
components of the mass density distribution cannot simply be superposed 
because of non-linear couplings amongst the components. 
Moreover the value $\rmsvel\sim30\kmps$ that they quote for the 
disk stars has little to do with the problem at hand, because the 
disk stars
are supported against the galactic gravity mainly by their circular 
motion
about the centre of the Galaxy. Also, in Ref. 4 of GKT, there 
is no attempt to fit the rotation curves to the actual data. 

In contrast with all earlier analyses, we have formulated the problem 
to directly
address the phase-space structure. There are two adjustable parameters 
in our
model, the central density $\rho_{\dm}(r=0)$ and the velocity dispersion 
$\rmsveldm$ of the dark matter particles. By fitting the rotation curve 
of the Galaxy up to $R\sim 20\kpc$, both the parameters were 
determined, even though we placed particular emphasis on the value
of $\rmsveldm$. The $\chi^2$ for $\rmsveldm$ = $300\kmps$ is more than 
4 times 
the value for $\rmsveldm = 600\kmps$, ruling out smaller dispersion 
velocities. 
   
In summary, we maintain that the conclusions 
reached in our Letter\cite{crb} are correct, robust and not in conflict 
with any established observational facts.  

\end{document}